# Non-equilibrium transport of inhomogeneous shale gas under unltratight confinement

Baochao Shan[1], Runxi Wang[1], Peng Wang[1], Yonghao Zhang[2*], Liehui Zhang[3], Zhaoli Guo[1*]

1. *State Key Laboratory of Coal Combustion, Huazhong University of Science and Technology, Wuhan, Hubei 430000, China*

2. *James Weir Fluids Laboratory, Department of Mechanical and Aerospace Engineering, University of Strathclyde, Glasgow G1 1XJ, UK*

3. *State Key Laboratory of Oil and Gas Reservoir Geology and Exploitation, Southwest Petroleum University, Chengdu, Sichuan 610500, China*

\* Corresponding authors: Zhang (*yonghao.zhang@strath.ac.uk*) and Guo (*zlguo@hust.edu.cn*)

**Abstract:** The non-equilibrium transport of inhomogeneous and dense gases highly confined by surface is encountered in many engineering applications. For example, in the shale gas production process, methane is extracted from ultra-tight pores under high pressure so the gas is inhomogeneous and dense. Currently, the complex non-equilibrium transport of inhomogeneous and dense gases where gas surface interactions play a key role is commonly investigated by molecular dynamics or on a continuum-assumption basis. Here, a tractable kinetic model based on the generalized Enskog equation and the mean-field theory is employed to couple the effects of the volume exclusion and the long-range intermolecular attraction forces. The interactions between gas molecules and confined surface are modelled by a $10-4-3$ Lennard-Jones potential, which can capture gas surface adsorption. The cross-sectional density profiles of methane under different confinements are in good agreement with the molecular dynamics results reported in the literature, and the transport behaviors are validated by the non-equilibrium molecular dynamics. The velocity of methane flow in shale matrix is plug-like due to its dense characteristics in nanopores. The influence of pressure, temperature, pore size and shale composition on density and velocity profiles is analyzed quantitatively. Our results show that the Klinkenberg correction is not applicable to model shale gas flow in the production process; the Navier-Stokes model using the second-order slip boundary condition cannot produce the proper velocity profiles, and consequently fails to predict the accurate flow rate in nanopores. This study sheds new light on understanding the physics of non-equilibrium dense gas flows in shale strata.

Key words: **shale gas reservoir; dense gas; the mean-field theory; adsorption; slip flow; nanopores**

**Introduction**

The unconventional shale gas reservoir is well recognized by its unique characteristics, such as ultra-tight strata, high pressure and temperature, multiple storage types and complex flow mechanisms (Zhang et al. 2019; Cristancho-Albarracin et al. 2017; Vasileiadis et al. 2018; Mehrabi et al. 2017; Darabi et al. 2012), which pose research and technological challenges for its efficient development. Although great success has been achieved for its commercial development as a result of technological advancement in horizontal drilling and multi-stage hydraulic fracturing (Zhao et al. 2018), the pore-scale understanding of gas transport mechanisms in shale gas production is still poor, which limits our ability to quantify long-term gas production (Nazari Moghaddam & Jamiolahmady 2016).

Unlike the conventional reservoirs, shale gas matrix is mainly composed of nanometer pores, which lead to large internal surface areas and provide enormous sites for gas storage (Li et al. 2018; Cai et al. 2019a). Since the flow in shale matrix is highly confined, the Knudsen number ($Kn$), defined as the ratio of gas mean free path $\lambda$ to a characteristic length $H$, e.g. pore diameter, can be very large (Nazari Moghaddam & Jamiolahmady 2016), ranging from 0.0003 to 3 under the typical shale gas reservoir conditions (Zhang et al. 2019). The rarefaction effect, usually interpreted as gas slippage and Knudsen diffusion at surfaces (Wang et al. 2018), is believed to play an important role in shale gas transport (Kazemi & Takbiri-Borujeni 2017; Cai et al. 2019b). So far, the rarefaction effect in shale gas transport has mainly been considered by gas kinetic theory for dilute gases. However, for the high pressure production process, shale gas is no longer a dilute gas, and the size of gas molecules cannot be ignored.

It is known that the Darcy's law, which is based on the Navier-Stokes (N-S) equation, works only in the continuum flow regime limit ($Kn < 0.001$). Thus, various slip boundary conditions have been proposed to account for the rarefaction effects when $0.001 < Kn < 0.1$. A comprehensive summary and comparison of these slip boundary conditions can be referred to our recent review article (Zhang et al. 2019). With further increase of the $Kn$, the Knudsen diffusion is introduced to account for the significant gas-solid interactions in nanoscale channels (Darabi et al. 2012). There has always been a controversy on the role of the Knudsen diffusion. Some adopt the Knudsen diffusion to describe flow behaviors in the free molecular flow regime when $Kn > 10$ (Rahmanian et al. 2013; Wu et al. 2016a), while others deem that the Knudsen diffusion induces the gas slippage near the rock surface (Ertekin et al. 1986; Florence et al. 2007). Javadpour (Javadpour 2009) proposed an apparent permeability model for fluid flow in shale nanopores, which is a linear superposition of slip flow and Knudsen diffusion. Based on this conception, a large number of apparent permeability models (Wu et al. 2016a; Huang et al. 2018) have been proposed to account for the viscous flow, slippage effect, Knudsen diffusion and even the surface diffusion in shale gas reservoirs. Although most of the models can fit the experimental data well, these empirical models rely on many unmeasurable parameters, which restrict their applicability in quantifying flow properties of shale. An improved model is requested to accurately describe gas/surface interactions and transport process at the pore scale.

The Boltzmann equation and its models can properly describe rarefaction effects, and are applicable to all the flow regimes from continuum to free molecular, provided that the gas is sufficiently dilute where the finite size of gas molecules can be ignored and only localized binary collisions need to be considered. Previously, the rarefied gas dynamics were commonly employed to account for the large $Kn$ effect in shale gas reservoirs according to the similarity criterion, where the Knudsen number was taken as the only criterion number (Nazari Moghaddam & Jamiolahmady 2016; Wu et al. 2016a; Bezyan et al. 2019; Shariati et al. 2019; Tan et al. 2020). However, the similarity between the microscale gas flow and the rarefied gas flow with the same $Kn$ only stands for a perfect gas (Wang 2003; Wang et al. 2008). It breaks down for shale gas, where the real gas effects must be considered due to the high pressure condition (Cai et al. 2018; Wu et al. 2017; Zhang et al. 2018). The Boltzmann equation needs to be extended to consider dense gas effect (Wu et al. 2016b). Enskog was the first to extend the Boltzmann equation to hard-spherical dense gases by considering the collisional transfer of momentum and the increased collision frequency due to the finite size of gas molecules. The nonlinear Enskog equation was revised by Beijeren and Ernst, known as the revised Enskog theory, which is not restricted to small spatial non-uniformities and gives results in agreement with irreversible thermodynamics (Beijeren & Ernst 1973). The long-range attractive force, which is not considered in the Enskog or revised Enskog theory, was taken into account by the Enskog-Vlasov equation using the mean-field theory (Karkheck & Stell 1981). In these extended Enskog-type theory, the collision operator is very complicated and the computational cost is formidable for practical applications, which needs further simplifications. The Enskog collision operator was expanded into a Taylor series of molecular diameter retaining up to the first order terms, where the zeroth-order term gave the Boltzmann collision operator and the velocity distribution function was approximated by the local equilibrium distribution function in the first-order terms. Luo (Luo 1998, 2000) derived a non-ideal gas lattice Boltzmann method (LBM) model directly from the Enskog equation for dense gases in the presence of an external force, where the Boltzmann collision operator in the Enskog equation was simplified by a Bhatnagar–Gross–Krook (BGK) collision operator. Similarly, an Enskog-Vlasov-type model was developed by He *et al.* (He & Doolen 2002; He et al. 1998) using the BGK approximation to the Boltzmann collision term in the Enskog equation. However, no wall potential was considered in their works, and the coupling of time step and grid size in the standard LBM as well as the limited discretized velocity numbers so these models can only be applied to unconfined continuum flows (Wang et al. 2019).

A tractable gas kinetic model was proposed by Guo et al. (Guo et al. 2005a) to predict the static structure and flow dynamics of confined fluids with strong inhomogeneity, where the finite size of fluid molecules, volume exclusion effect, long-range molecular interaction and pore confinement effects were simultaneously considered. The model was employed to study the temperature dependence of the velocity slip (Guo et al. 2005b) and proved to be in qualitatively agreement with molecular dynamics simulations (Guo et al. 2006b). Later, a generalized hydrodynamic model was derived for fluid flows from nanoscale to macroscale, which captured the strong inhomogeneity of fluid structures at the nanometer scale and degenerated to the conventional N-S equation at the macroscale (Guo et al. 2006a). However, non-equilibrium effects were not properly considered. Recently, the discrete unified gas kinetic scheme (DUGKS) was employed to solve the kinetic model (Guo et al. 2005a) of the dense fluid system with strong inhomogeneity (Shan et al. 2020), which captured the dense gas effect properly, but only the wetting cases of the surfaces were considered.

In this paper, employing the tractable kinetic model (Guo et al. 2005a), the inhomogeneity and non-equilibrium shale gas transport behaviors are studied for both the wetting and non-wetting cases, where the pore confinement effect, gas adsorption behaviors and the dense gas effects under high pressure and high temperature conditions are simultaneously taken into account in a self-consistent way. Without any arbitrary empirical parameter, this work may serve as a powerful simulation tool to systematically study the gas/surface interactions and transport in shale gas reservoirs.

**1 Flow characteristics in shale matrix**

Shale matrix is mainly composed of organic matter and inorganic matter with a significant number of nanopores. Due to the small pore size, the Knudsen number can be high in nanopores (Wang et al. 2018), where the rarefaction effects need to be considered. The Knudsen number for non-ideal gas can be calculated by (Chapman & Cowling 1970)

$$Kn = \frac{1}{\sqrt{2}n\pi\sigma^2 \chi H}, \quad (1)$$

where $n$ is the number density, $\sigma$ is the molecular diameter, $\chi$ is the radial distribution function, and $H$ is the characteristic length of the flow path.

Due to the high pressure and high temperature conditions, shale gas transport in the highly-confined pores needs to consider the dense gas effect and confinement effect. Table 1 clarifies gas flow systems (Bird 1994; Gad-el-Hak 1999), including (1) the dilute gas limit $\delta / \sigma$: the ratio of the average distance between molecules $\delta$ ($\delta = n^{-1/3}$) to the molecule size $\sigma$; (2) the pore confinement limit $H / \sigma$: the ratio of the characteristic length $H$ to the molecular size $\sigma$; (3) the statistical fluctuation limit $H / \delta$: the ratio of the characteristic length $H$ to the average distance between molecules $\delta$.

Table 1: The criterion numbers and their limits for gas system determination (Bird 1994)

| Dilute gas limit | Pore confinement limit | Continuum limit | Statistical fluctuation limit |
|---|---|---|---|
| $\delta/\sigma = 7$ | $H/\sigma = 38$ | $Kn = \lambda/H = 0.1$ | $H/\delta = 100$ |

According to Table 1, the effective limits and the typical shale gas flow domain are shown in Figure 1, where the shale gas components are taken from Javadpour et al. (Javadpour et al. 2007) at the temperature of 323 K. These limits can be explained specifically as follows:

(1) The dilute gas limit: when $\delta/\sigma < 7$, the gas molecular size is comparable to the average distance between molecules $\delta$, and it can no longer be ignored. Under this circumstance, the dilute gas assumption for the Boltzmann equation is not valid. For a typical shale gas reservoir (Javadpour et al. 2007), the dense gas effect becomes prominent when pressure is larger than 0.2 MPa, as shown in Figure 1. Thus, all shale gas reservoirs are dense gas systems, which cannot be described by the Boltzmann equation. It is imperative to take the finite size of gas molecules and their non-local collisions into account simultaneously to describe gas dynamics.

(2) The pore confinement limit (Wu & Chen 2016): when $H/\sigma < 38$, the pore confinement effect should be considered, which corresponds to a pore diameter $H \approx$ 15.4 nm. Therefore, confinement effect is significant in shale gas transport at the pore scale.

(3) The continuum limit: the N-S equations can be adopted with a no-slip boundary condition when $Kn < 0.001$ in the continuum regime, or with a slip boundary condition when $0.001 < Kn < 0.1$ in the slip flow regime. It breaks down for a confined dense fluid system, where the gas properties become inhomogeneous over a molecular dimension.

(4) The statistical fluctuation limit: this is the limit for statistical fluctuations when calculating macro variables from the microscopic information. When $H/\delta > 100$, there is sufficient microscopic information to calculate the macro variables. However, the statistical fluctuation becomes significant with the decrease of the pore size, which increases the complexity to study the gas flow behaviors in shale nanopores.

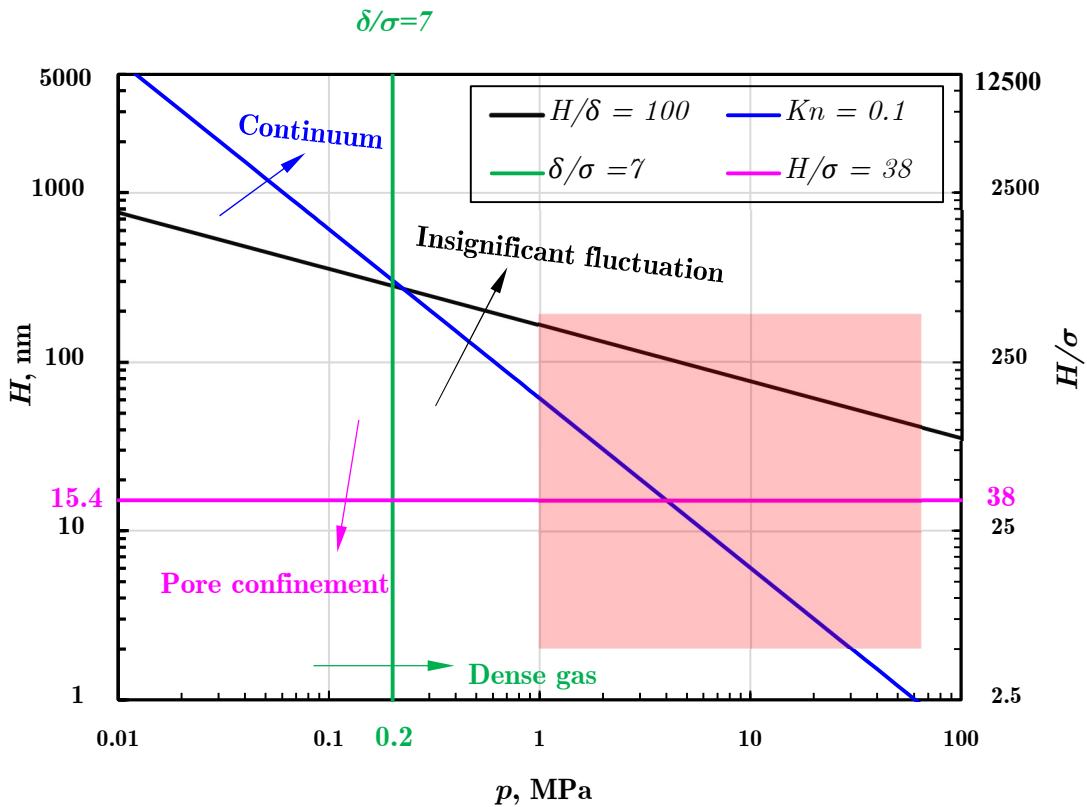

Figure 1: Effective limits of different flow models revised from Bird (Bird 1994). The typical shale gas flow

domain is displayed in the shadow region for pores ranging from 2 nm to 200 nm and pressure ranging from 1MPa to 60 MPa at the temperature of 323 K. The shale gas components are taken from Javadpour et al. (Javadpour et al. 2007), where the $CH_4$ accounts for 87.4%, $C_2H_6$ accounts for 0.12% and $CO_2$ accounts for 12.48%.

**2 Model establishment and solution**

In this section, the physical background, simplified conceptual model and the tractable kinetic model for non-equilibrium transport of shale gas under confinement are briefly introduced. As we know, both organic and inorganic pores are very developed in shale matrix (Figure 2). These pores provide adsorption sites for gas storage (Li et al. 2016) and flow paths for gas transport. Although different flow mechanisms occur in organic and inorganic pores (Zhang et al. 2017), it will be shown that our model can describe flow dynamics in both pore types, with different energy parameters of pore surface and solid density to control the wettability, which is consistent with techniques in MD simulations to model organic and inorganic materials.

**2.1 Physical background**

The sketch of inorganic pores and organic pores of Longmaxi shale formation, ranging from several nanometers to several hundred nanometers, can be observed directly by the Scanning Electron Microscopy (SEM) technique, as shown in Figure 2. Although the porosity in inorganic matrix (Figure 2a) is much lower than that in organic kerogen (Figure 2b), the storage and transport of shale gas in inorganic pores can still not be ignored.

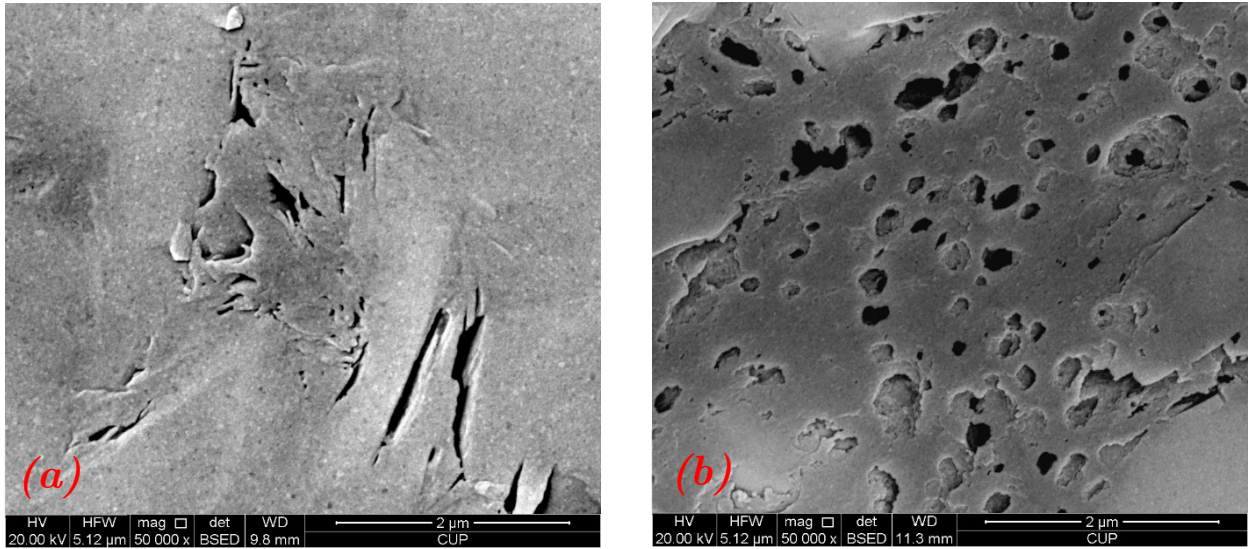

Figure 2: The SEM images of polished shale samples in Longmaxi shale formation: (*a*) inorganic pores and (*b*) organic pores. The pores are shown in black.

**2.2 Conceptual model**

In shale nanopores, the finite size of gas molecules, gas molecular interactions and pore confinement effects can no longer be ignored. The practical pore shapes in shale matrix are very complex, as shown in Figure 2. In this paper, the nanopores are simplified by two paralleled plates with different separations. As shown in Figure 3, gas molecule size is comparable to the separation of the channel, which becomes a key parameter in controlling the flow behaviors at the nanometer scale. Meanwhile, the wall potential affects the static structure of gas molecules in nanopores, forming one or more adsorption layers near the solid surface. As we can see from Figure 3*a*, gas molecules fill in the small pores due to the strong solid-gas interactions superposed by the top and bottom surfaces, which is in accordance with the micro pore filling theory for gas adsorption. With the increase of wall separation, the effects of the surfaces on fluid molecules decrease, and a second adsorption layer may occur at the center of the flow region, as shown in Figure 3*c*. There is still no bulk region under this condition. There will be several adsorption layers near the wall and a bulk region at the center when the flow path is large enough (Figure 3*c*). All three cases (Figure 3*b, c* and *d*) imply that there may be strong inhomogeneity for gas distribution across the confinement in shale nanopores. The different flow behaviors in organic pores and inorganic pores are resulted from the difference of pore characteristics, such as the energy parameter. Therefore, it is necessary to take the gas and solid characteristics into account in the mathematical model.

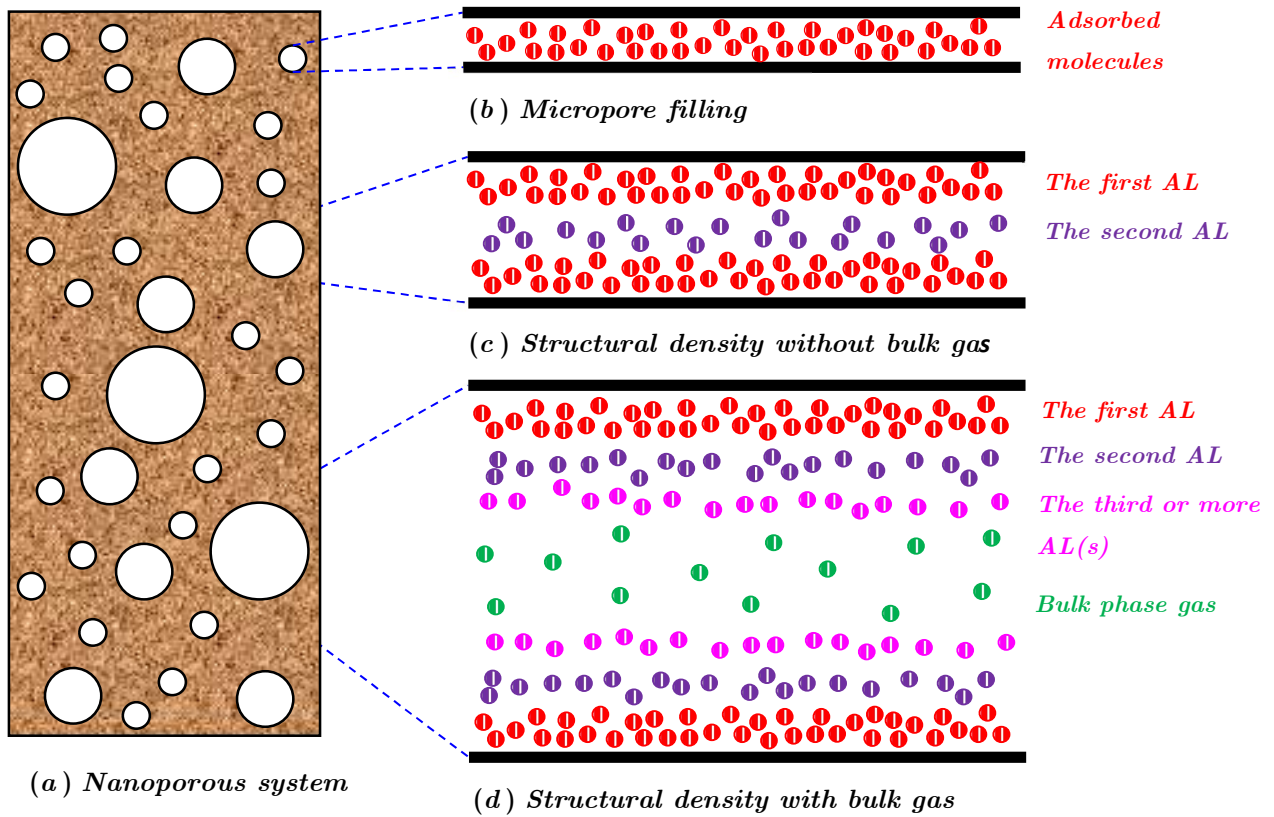

Figure 3: Schematic of pores sizes distribution in shale matrix (*a*) as well as the gas molecules distribution in pores of different sizes (*b*), (*c*), and (*d*). AL is short for adsorption layer for convenience.

**2.3 Kinetic model for dense gases**

Here, a tractable kinetic model proposed by (Guo et al. 2005a) is employed to study shale gas density distribution and non-equilibrium dynamics in nanopores. The governing equation can be described by

$$\partial_t f + \boldsymbol{\xi} \cdot \boldsymbol{\nabla}_r f - a \cdot \boldsymbol{\nabla}_\xi f = \Omega, \tag{2}$$

where $f(\boldsymbol{r}, \boldsymbol{\xi}, t)$ is the velocity distribution function of molecular velocity $\boldsymbol{\xi}$ at spatial position $\boldsymbol{r}$ and time $t$, $\partial_t$ represents partial derivative in terms of time $t$, $\boldsymbol{\nabla}_r$ and $\boldsymbol{\nabla}_\xi$ represent gradient operators in terms of space $\boldsymbol{r}$ and velocity $\boldsymbol{\xi}$, respectively; $\Omega$ is the extended Enskog collision operator, and $a$ is the force acceleration term expressed as

$$a = m^{-1} \boldsymbol{\nabla}_r \left( \phi_e + \phi_m \right), \tag{3}$$

where $m$ is the molecular mass, $\phi_e$ is the external potential, including the wall potential and the driving force related potential; and $\phi_m$ is related to the attractive potential, which is modeled according to the mean-field theory as (He et al. 1998)

$$\phi_m(\boldsymbol{r}) = \int_{|\boldsymbol{r}'|>\sigma} n(\boldsymbol{r} + \boldsymbol{r}') \phi_a(|\boldsymbol{r}'|) d\boldsymbol{r}', \tag{4}$$

where $\phi_a$ is the attractive part of the Lennard-Jones (LJ) fluid potential

$$\phi_a(\boldsymbol{r}) = 4\varepsilon_{ff} \left[ \left( \frac{\sigma_{ff}}{\boldsymbol{r}} \right)^{12} - \left( \frac{\sigma_{ff}}{\boldsymbol{r}} \right)^{6} \right], \tag{5}$$

where $\varepsilon_{ff}$ and $\sigma_{ff}$ are the energy and range parameters of fluid-fluid interactions, respectively.

According to the projection method in the revised Enskog theory for homogeneous hard-sphere fluids (Dufty et al. 1996), the extended Enskog collision operator can be projected into a Boltzmann collision operator $\Omega_B$ and an excess collision operator $\Omega_E$. Considering the fact that shale gas production can be seen as an isothermal process, the Boltzmann collision operator can be further simplified into the Bhatnagar-Gross-Krook type operator as

$$\Omega_B = -\frac{1}{\tau} \left[ f - f^{(eq)} \right], \tag{6}$$

where $\tau$ is the relaxation time, and $f^{(eq)}$ is the local equilibrium distribution function

$$f^{eq} = n\left(\frac{m}{2\pi k_B T}\right)^{3/2} \exp\left[-m\frac{(\boldsymbol{\xi}-\boldsymbol{u})^2}{2k_B T}\right], \tag{7}$$

where $k_B$ is the Boltzmann constant, $T$ is the system temperature, and $\boldsymbol{u}$ is the flow velocity.

Meanwhile, the excess collision operator $\Omega_E$ can be expressed as

$$\Omega_E = -V_0 f^{(eq)}(\boldsymbol{\xi}-\boldsymbol{u})\cdot\left[2\boldsymbol{A}\chi(\bar{n}) + \boldsymbol{B}\bar{n}\right], \tag{8}$$

where $V_0$ is related to molecular diameter $\sigma$ by $V_0 = 2\pi\sigma^3/3$, $\bar{n} = \int w(\boldsymbol{r}')n(\boldsymbol{r}+\boldsymbol{r}')d\boldsymbol{r}'$ is the local average density with $w(\boldsymbol{r})$ being a weighting function (Tarazona 1985; Vanderlick et al. 1989), which was commonly used in free energy density functional theory (DFT) to study inhomogeneous fluid systems; $\chi$ is the radial distribution function (RDF) for homogenous hard-sphere fluid (Carnahan & Starling 1969). Here, the RDF $\chi$ is evaluated at the local average density, rather than the local density, to account for the possible inhomogeneity of shale gas in nanopores, as mentioned in Figure 3. The two terms $\boldsymbol{A}$ and $\boldsymbol{B}$ are

$$\boldsymbol{A} = \frac{1}{D}\int_{|\boldsymbol{r}'|<\sigma/2} \boldsymbol{r}'\bar{n}(\boldsymbol{r}+\boldsymbol{r}')d\boldsymbol{r}', \tag{9}$$

and

$$\boldsymbol{B} = \frac{1}{D}\int_{|\boldsymbol{r}'|<\sigma/2} \boldsymbol{r}'\chi\left[\bar{n}(\boldsymbol{r}+\boldsymbol{r}')\right]d\boldsymbol{r}', \tag{10}$$

where $D = \pi\sigma^5/120$.

The surface potential can be calculated from the integration of a continuous distribution of all the solid atoms, which interact with gas molecules through the 12 – 6 L-J potential. According to Steele (Steele 1973), the following 10 – 4 – 3 wall potential is obtained after the integration

$$\phi_w(z) = 2\pi n_s \varepsilon_{wf}\left[\frac{2}{5}\left(\frac{\sigma_{wf}}{z}\right)^{10} - \left(\frac{\sigma_{wf}}{z}\right)^4 - \frac{\sigma_{wf}^4}{3\Delta(z+0.61\Delta)^3}\right], \quad \Delta = \sigma_{wf}/\sqrt{2}, \tag{11}$$

where $n_s$ is the solid molecular density, $\varepsilon_{wf}$ and $\sigma_{wf}$ are energy and range parameters of gas-surface interactions, respectively; and $z$ is the normal distance of a gas molecular from the shale surface. According to the superposition theory of the potential, a fluid molecule in the nano-slit is under the action of the top and bottom walls simultaneously, *i.e.*,

$$\phi_e = \phi_w(z) + \phi_w(H-z), \ 0 \leq z \leq H. \tag{12}$$

The effective molecular diameter $\sigma$ can be calculated by (Cotterman et al. 1986)

$$\sigma \approx \frac{1 + a_1 T_r}{1 + a_2 T_r + a_3 T_r^2}, \tag{13}$$

where $a_1 = 0.2977$, $a_2 = 0.33163$, $a_3 = 0.00104771$; and $T_r$ is the reduced temperature expressed as

$$T_r = \frac{k_B T}{\varepsilon}. \tag{14}$$

The range and energy parameter of gas-surface interactions $\sigma_{wf}$ and $\varepsilon_{wf}$ can be obtained from those of gas-gas interactions $\sigma_{ff}$ and $\varepsilon_{ff}$, and those of surface-surface interactions $\sigma_{ww}$ and $\varepsilon_{ww}$ according to the Lorentz-Berthelot combination rule (Morciano et al. 2017)

$$\sigma_{wf} = \frac{\sigma_{ww} + \sigma_{ff}}{2}, \ \varepsilon_{wf} = \sqrt{\varepsilon_{ww} \varepsilon_{ff}}. \tag{15}$$

The number density $n$ and flow velocity $\boldsymbol{u}$ are moments of the distribution function, they can respectively calculated by

$$n = \int f d\boldsymbol{\xi}, \tag{16}$$

and

$$\boldsymbol{u} = n^{-1} \int \boldsymbol{\xi} f d\boldsymbol{\xi}. \tag{17}$$

In solving the kinetic model, the relaxation time $\tau$ in Eq.(6) is determined by (Bitsanis et al. 1987)

$$\tau = \frac{\mu(\overline{n})}{nk_BT}, \tag{18}$$

where $\mu(n)$ is the viscosity for a homogeneous dense fluid with density $n$, which is (Guo et al. 2006a; Chapman & Cowling 1970)

$$\mu(n) = \mu_0 n V_0 \left(Y^{-1} + 0.8 + 0.7614Y\right), \tag{19}$$

with

$$\mu_0 = \frac{5.0}{16\sigma^2}\sqrt{\frac{k_BT}{\pi}},\ Y = nV_0\chi(n),\ \chi(n) = \frac{1-0.5\eta}{(1-\eta)^3},\ \eta = \frac{nV_0}{4},\ V_0 = \frac{2\pi\sigma^3}{3}. \tag{20}$$

**2.4 Model solution**

The kinetic equation (2), which governs the non-equilibrium gas dynamics in shale strata, is solved by the DUGKS (Guo et al. 2013), which utilizes the advantages of the lattice Boltzmann method (LBM) (McNamara & Zanetti 1988) and the unified gas kinetic scheme (UGKS) (Xu & Huang 2010). The discrete unified gas kinetic scheme (DUGKS) is capable of simulating flows at all Knudsen numbers in rarefied gas dynamics. Recently, the DUGKS has been applied to solve the strongly inhomogeneous fluid system is also proved to be practicable. Readers can refer to Shan et al. (Shan et al. 2020) for the details of the DUGKS. In this paper, the bounce-back rule, together with the surface-fluid interaction of Eqs.(11) and (12), constitutes the boundary condition, where the slip phenomenon can be modeled by the bounce-back scheme and the adsorption layers are formed due to the surface-fluid interaction.

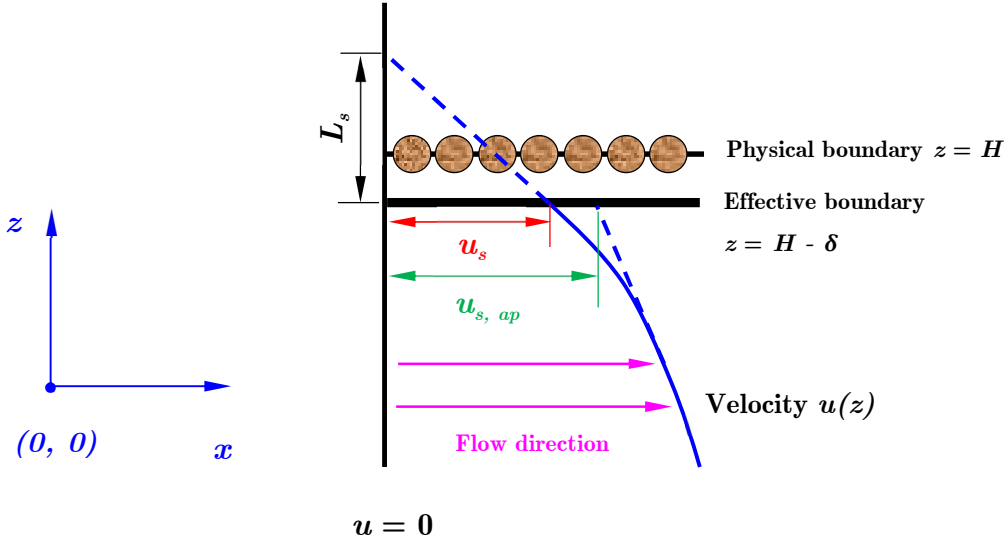

Figure 4: Schematic of slip phenomenon in shale nanopores, in which the physical boundary and effective boundary do not coincide. The $u_s$ is the true slip velocity at the effective boundary and the $u_{s,\,ap}$ is the apparent slip velocity, which is obtained from the extrapolation of macroscopic velocity distribution. $\delta$ is the distance between the effective boundary and the physical boundary condition, which is taken at the mass center of the solid molecules.

For the nano-confined dense gas, a void-region, shown as the region between the physical boundary and effective boundary in Figure 4, exists between the shale surface and the first adsorbed fluid layer due to the surface-fluid and fluid-fluid interactions. The fluid molecules, which are in contact with the effective boundary, move forward relative to the shale surface due to the lattice structure and the chemical properties of the surface (Bhadauria et al. 2015). This is the so-called slip phenomenon characterized by a slip velocity $u_s$. The slip velocity $u_s$ can be coupled into the bounce-back rule as

$$f\left(\boldsymbol{\xi}_i\right) = f\left(-\boldsymbol{\xi}_i\right) + 2\rho_r\, W_i\, m\, \frac{\boldsymbol{\xi}_i \cdot \boldsymbol{u}_s}{k_B T},\quad \boldsymbol{\xi}_i \cdot \boldsymbol{n} > 0, \tag{21}$$

where $\boldsymbol{n}$ is the unit vector normal to the rock surface pointing to the flow domain, $W_i$ is

related to $\omega_i$ by

$$W_i = \omega_i \left(\frac{m}{2\pi k_B T}\right)^{3/2} \exp\left(-m\frac{|\boldsymbol{\xi}_i|^2}{2k_B T}\right), \tag{22}$$

where $\omega_i$ is the weight coefficients of the numerical quadrature at the discrete velocity $\boldsymbol{\xi}_i$, which can be adsorbed into the distribution functions (Guo et al. 2013); and the fluid density at the rock surface $\rho_r$ is determined by

$$\rho_r = \left[\sum_{\boldsymbol{\xi}_i\cdot\boldsymbol{n}=0} f(\boldsymbol{\xi}_i) + 2\sum_{\boldsymbol{\xi}_i\cdot\boldsymbol{n}<0} f(\boldsymbol{\xi}_i)\right] \bigg/ \left(1 - \frac{2m}{k_B T}\sum_{\boldsymbol{\xi}_i\cdot\boldsymbol{n}>0} W_i \boldsymbol{\xi}_i \cdot \boldsymbol{u}_s\right). \tag{23}$$

Note that the $u_s$ is the true slip velocity, which depends on the physical nature of gas-solid interactions, as shown in Figure 4. The apparent slip velocity $u_{s,\,ap}$ is extrapolated from the macro velocity, which is usually larger than the true slip velocity $u_s$. According to Bhadauria et al. (Bhadauria et al. 2015), the true slip velocity can be expressed by

$$u_s = \frac{G_x}{\zeta_0} \int_{-H/2}^{0} n(z) dz = H \frac{G_x}{2\zeta_0} n_0, \tag{24}$$

where $G_x$ is the driving force in the flow direction, and $\zeta_0$ is the friction coefficient. The friction coefficient can be determined by the generalized Langevin equation(Bhadauria et al. 2015), MD simulation (Bocquet & Barrat 1994) or simply by fitting the experimental data. The value of friction coefficient $\zeta_0$ is fixed once the thermodynamic state of the fluid and the physical properties are specified. Therefore, the friction coefficient $\zeta_0$ only needs to be determined once for a specific thermodynamic case, after which it can be used to calculate the slip velocity $u_s$ according to Eq.(24) for other cases with different driving forces or channel widths.

**2.5 Model validation**

Molecular dynamics (MD) simulation results from the previous literature (Somers & Davis 1992) are collected to validate the static structures of fluid confined between two paralleled plates at the nanometer scale. Then, the non-equilibrium molecular dynamics (NEMD) is performed to validate both the static and flow behaviors of methane in graphene slits. In all of our simulations, the mesh size in the $z$ direction is set to be $\Delta z = 0.01\ H$, which is sufficient to obtain the grid independent results. A 8×8 Gauss-Hermit discrete velocity set distributed in $[-4\sqrt{2k_B T/m},\ 4\sqrt{2k_B T/m}]$ is adopted in the $x$ and $z$ direction, respectively. Meanwhile, the Courant-Friedrichs-Lewy (CFL) number is chosen as 0.2.

**2.5.1 Static structures**

The cross-sectional density distributions with wall separations ranging from $2\sigma$ to $8\sigma$ are compared in Figure 5, where the 10 – 4 – 3 LJ potential is employed for fluid-surface interactions. The system is maintained at a reduced temperature of $T_r = k_B T/\varepsilon_{ff} = 1.2$, and the energy parameters for solid-solid interactions $\varepsilon_{ww}$ equals to that for fluid-fluid interactions $\varepsilon_{ff}$. The pore averaged density for different cases is displayed in Table 2. As we can see from Figure 5, the static structures of fluids confined in nanometer spaces are in quantitatively good agreement with MD results, which shows the accuracy of our method in capturing the structural inhomogeneity of dense gases.

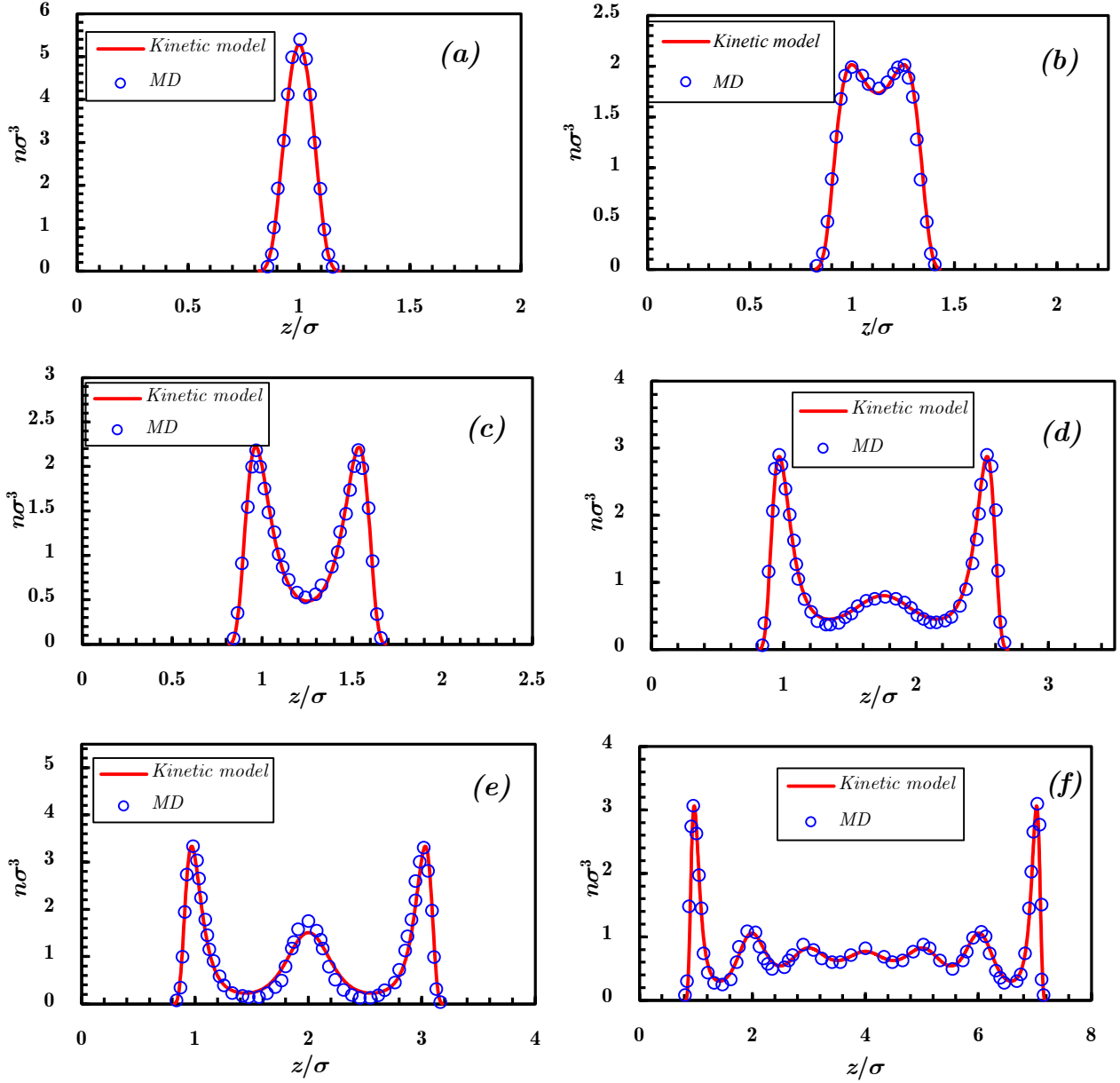

Figure 5: Comparison of our kinetic results with the MD data for density distribution across the channel. The fluids are confined between two paralleled plates at the temperature of $T_r = 1.2\ \varepsilon_{ff}\,/\,k_B$. The pore average density for different pores is shown in Table 2. The MD results are from Somers & Davis (Somers & Davis 1992)

Table 2: Simulation parameters for static structure calculation

| Figure 5 | a | b | c | d | e | f |
|---|---|---|---|---|---|---|
| Pore width, $\sigma$ | 2.0 | 2.25 | 2.5 | 3.5 | 4.0 | 8.0 |
| Average density, $\sigma^{-3}$ | 0.414 | 0.368 | 0.369 | 0.508 | 0.565 | 0.625 |

**2.5.2 Transport behaviors**

In this section, the non-equilibrium molecular dynamics (NEMD) simulation is performed to simulate the transport behavior of methane in nano-confined graphene channels, which are commonly used to represent shale gas flow in organic nanopores. The LAMMPS package is employed to perform the NEMD simulation, where a driving force of $1.0 \times 10^{-4}$ kcal/(mol Å) is exerted on methane molecules in the $x$ direction to mimic a pressure driven flow. In the simulation, the temperature is taken as 363 K; the pore averaged density, defined as $\rho_0 = \int_0^H \rho(z) dz / H$, is constrained at 190 kg/m$^3$. The schematic of organic nano-channel can be seen in Figure 6, where the shale rock is represented by the six-layer graphene, with the lateral extent of 55.38 × 56.57 Å$^2$ consisting of 1196 graphene atoms in each layer. The interlayer spacing is 3.41 Å. The periodical boundary conditions are employed in the $x$ and $y$ directions.

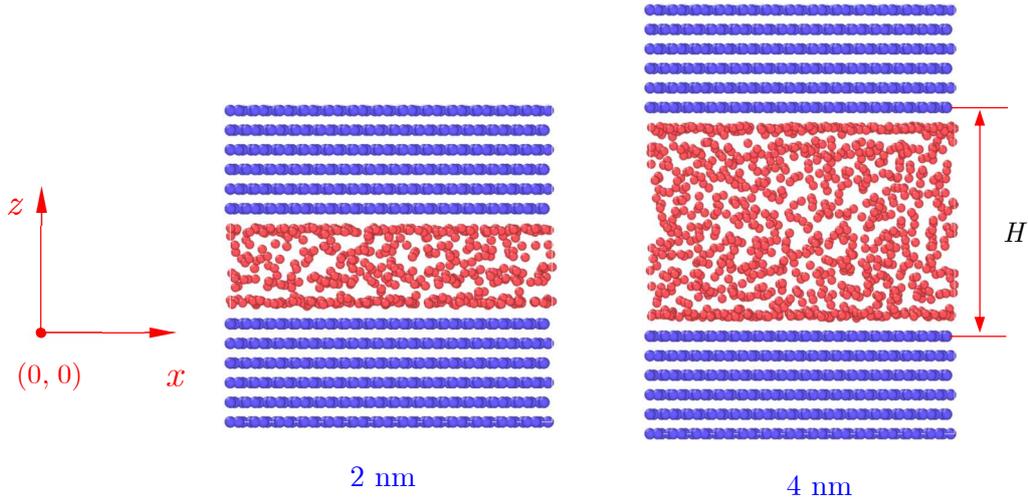

Figure 6: Snapshots of MD simulation box for 2 nm and 4 nm graphene channels, respectively; the organic shale rocks are represented by the top and bottom six-layer graphene, with the interlayer distance of 3.41 Å; the lateral extent of each graphene layer is 55.38 × 56.57 Å$^2$, consisting of 1196 graphene atoms.

Table 3: Parameters of graphene and methane used for the MD simulation

| Atom type | Mass (kg/mol) | $\varepsilon$ (kcal/mol) | $\sigma$ (nm) |
| --- | --- | --- | --- |
| Graphene carbon | 12.011 | 0.0700 | 0.355 |
| Methane | 16.040 | 0.2940 | 0.373 |

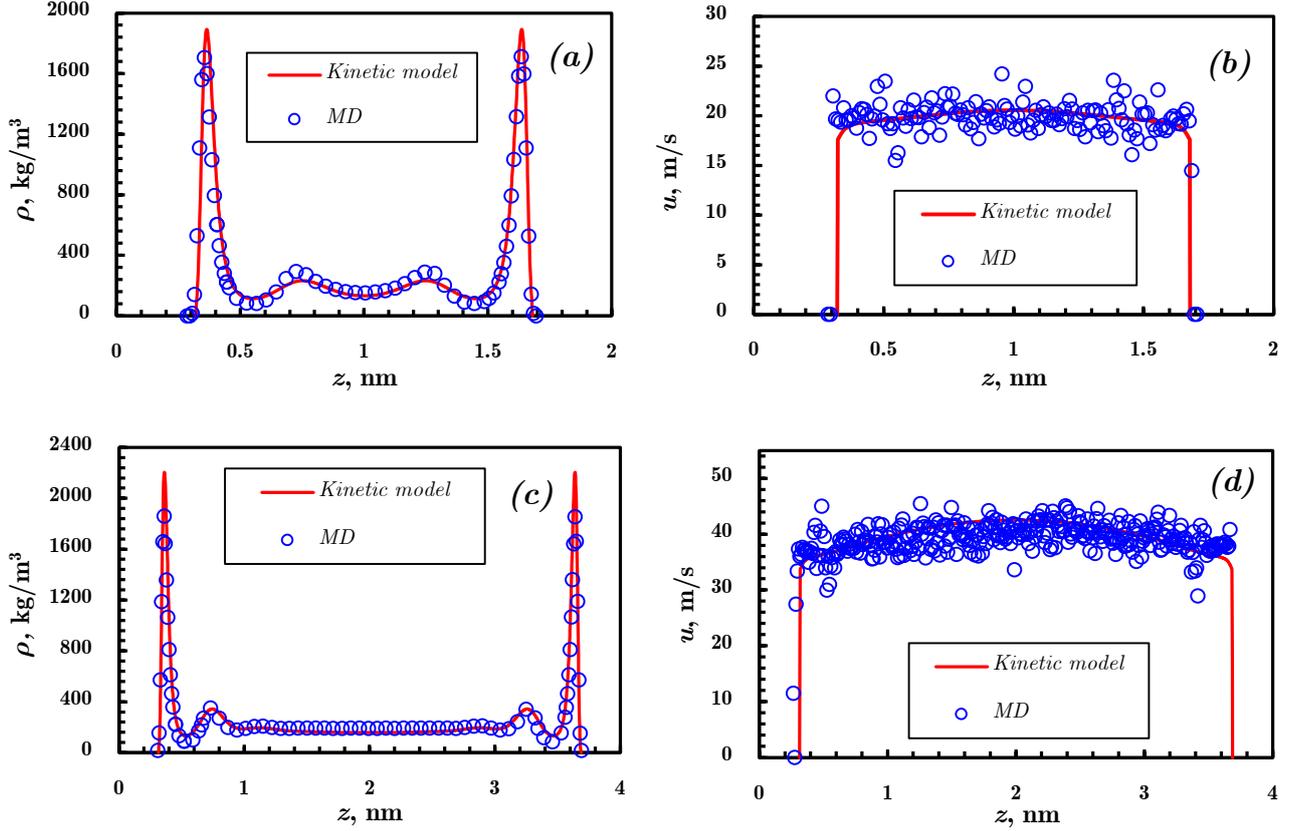

Figure 7: Comparison of the density and velocity profiles obtained from the MD simulations and the solution of kinetic model for shale gas flow in 2 nm (*a* and *b*) and 4 nm (*c* and *d*) slits. The organic slits are modeled by the perfect graphene, which is smooth and causes significant slip at the surface.

As shown in Figure 7, the density and velocity profiles of the kinetic model and NEMD simulations agree well with each other. No obvious bulk region is observed in the small nano-slit (2 nm), while it becomes very obvious when the slit increases to 4 nm. The slip velocity $u_s$ from the kinetic model is determined from the MD simulation for the 2-nm slit, from which the friction coefficient is calculated as $\zeta_0 = 1.94326$ (kJ ps)/(mol nm$^2$). The calculated friction coefficient $\zeta_0$ along with the Eq.(24) is employed to simulate the methane flow in a 4-nm silt, and the velocity profile is in good quantitative agreement with the MD simulations. Therefore, only one MD simulation is needed to determine the friction coefficient, and the kinetic model can then simulate the flow behaviors independently for different driving forces or channel widths.

## 3 Results and discussion

In this section, we report a comprehensive study of shale gas flow behavior in nanopores. The same set up for the NEMD simulations is used, and all the variables are expressed in the LJ units, as shown in Table 4. The surface roughness, which plays an important role in shale gas transport in nanopores, is ignored in our simulations. Generally, the rougher the rock surface, the severer the friction between the rock and the fluids, and the larger the friction coefficient. Thus, the friction coefficient is taken as twice of that obtained from the previous analysis to account for the effect of surface roughness.

Table 4: The kinetic model parameters for modeling methane flow in organic graphene at the temperature of $T$ = 363 K and the pore averaged density $n_0$ = 190 kg / m³. The width $H$ = 8.04 $\sigma$ represents a separation diameter of 3 nm.

| $H$, $\sigma_{ff}$ | $T$, $\varepsilon_{ff} / k_B$ | $n_0$, $\sigma^{-3}$ | $n_s$, $\sigma^{-3}$ | $G_x$, $\varepsilon_{ff} / \sigma$ |
|---|---|---|---|---|
| 8.04 | 2.45 | 0.371 | 5.810 | 0.00127 |

### 3.1 Effect of density

At a constant temperature, the relationship between density and pressure can be expressed by the equation of state. Figure 8 shows the static structure and dynamic behavior of methane flow in organic slits, with the pore averaged density $n_0$ = 0.1855 $\sigma^{-3}$, $n_0$ = 0.3710 $\sigma^{-3}$ and $n_0$ = 0.5565 $\sigma^{-3}$, respectively. According the NIST Chemistry WebBook, the selected pore average densities correspond to the pressure of $p$ = 16.7 MPa, $p$ = 37.9 MPa and $p$ = 83.2 MPa, respectively, at the temperature of 363 K. Clearly, both the adsorbed and bulk gas molecules increase with the pressure, but the density profiles fluctuate more under a higher pressure. The second adsorption layer is not obvious when the pore averaged density is 0.1855 $\sigma^{-3}$ (16.7 MPa), while the second adsorption layer is very obvious when the pore average density increases to 0.3710 $\sigma^{-3}$ (37.9 MPa). With further increment of the pore average density, a weak third adsorption layer occurs, as shown in Figure 8a when $n_0 = 0.5565\ \sigma^{-3}$ ($p = 83.2$ MPa). Thus, the adsorption mechanism of shale gas is affected by the pressure. Both the monolayer adsorption and multilayer adsorption may occur in shale gas reservoirs, depending on pressure. On the other hand, the velocity is also significantly influenced by the pore averaged density or pressure. With the same pressure gradient driving the flow, shale gas molecules move faster under a higher pressure. This also contributes to high gas production at the early stage, as the pressure is higher.

Similar to the Darcy's law, the apparent permeability $k_{app}$ can be calculated by $k_{app} = -Q_{app}\mu / (\rho \nabla p)$, where $Q_{app}$ is the apparent flow flux, which refers to the flux calculated from the kinetic model $Q_k$ or the conventional N-S equations with the slip boundary conditions $Q_s$, the flux $Q_s$ is equivalent to the Klinkenberg correction, and $\mu$ is the fluid viscosity evaluated at the pore averaged density $n_0$ according to Eq.(22). The parameters to calculate the apparent permeability with different densities are shown in Table 5. We also define the enhancement factor here, i.e. $\beta$ as $\beta = Q_{app} / Q_{ins}$, where $Q_{ins}$ is the intrinsic flow flux calculated from the Darcy equation without Klinkenberg correction. Different permeability and the enhancement factors are shown in Table 6.

Table 5: Parameters for calculating apparent permeability under different density conditions

| $n_0$, $\sigma^{-3}$ | $\mu$, $(m\varepsilon)^{0.5}/\sigma^3$ | $Kn$ | $Q_{ins}$, $(\varepsilon/m)^{0.5}\sigma^{-3}$ | $Q_k$, $(\varepsilon/m)^{0.5}\sigma^{-3}$ | $Q_s$, $(\varepsilon/m)^{0.5}\sigma^{-3}$ |
|---|---|---|---|---|---|
| 0.1855 | 0.172738 | 0.116750 | 0.058972 | 0.031559 | 0.100281580 |
| 0.3710 | 0.279415 | 0.043722 | 0.072914 | 0.129290 | 0.092041919 |
| 0.5560 | 0.536874 | 0.020954 | 0.056922 | 0.291867 | 0.064078688 |

Table 6: Permeability from different models and the enhancement factors: $k_k$ and $k_s$ are the permeability calculated from our kinetic model and the conventional N-S equations (with the slip boundary condition), respectively; $\beta_k$ and $\beta_s$ are the enhancement factor calculated from our kinetic model and the conventional N-S equations (with the slip boundary condition), respectively.

| $n_0$, $\sigma^{-3}$ | $k_{ins}$, $\sigma^2$ | $k_k$, $\sigma^2$ | $k_s$, $\sigma^2$ | $\beta_k$ | $\beta_s$ |
|---|---|---|---|---|---|
| 0.1855 | 43.30811 | 23.17625 | 73.6454443 | 0.535148 | 1.700500 |
| 0.371 | 43.30812 | 76.79308 | 54.6692154 | 1.77318 | 1.262332 |
| 0.556 | 43.34707 | 222.2607 | 48.7968763 | 5.127468 | 1.125725 |

As we can see from Table 6, the intrinsic permeability $k_{ins}$ is independent of pressure as an intrinsic property of the media. With pressure (*i.e.*, density in this case) decreasing, the apparent permeability from our kinetic model $k_k$ continuously decreases. This finding is totally different from the famous Klinkenberg effect, which believes that gas molecular moves faster under low pressure conditions due to the rarefaction effects. However, the gas is dense and flow is highly confined, the complex fluid/fluid/surface interactions cannot be described by gas kinetic theory of dilute gas. Further study is required to reveal underlying mechanisms.

The Klinkenberg modification relates the permeability with the pressure by $k_{app} = k_{ins}\left(1 + b_k / \overline{p}\right)$, where $b_k$ is the slippage factor. For fluid flow in slits, the slippage factor can be expressed as $b_k = 6c_k \lambda \overline{p} / H$, where $c_k$ is a fitting parameter approximately equaling to unity. Thus, the Klinkenberg correction can be written as $k_{app} = k_{ins}\left(1 + 6c_k Kn\right)$, which is equivalent to flow with the first order slip boundary condition to account for the rarefaction effects. As in Table 6, the apparent permeability from the Klinkenberg correction $k_{sp}$ increases with the decreasing pressure. This contradictory may arise from that the Klinkenberg correction is only applicable to dilute gas system, while shale gas is a dense one. In the derivation of Klinkenberg (Klinkenberg 1941), the low pressure condition and ideal gas equation of state are adopted, which are not applicable to shale gas system. Besides, the competition of fluid-fluid and fluid-rock interactions is ignored in the Klinkenberg correction (Klinkenberg 1941), while the shale gas system belongs to the high pressure system and fluid-fluid and fluid-rock interactions play important roles in fluid dynamics. The conventional Klinkenberg correction may underestimate the flux under high pressure conditions and overestimate the flux under low pressure conditions.

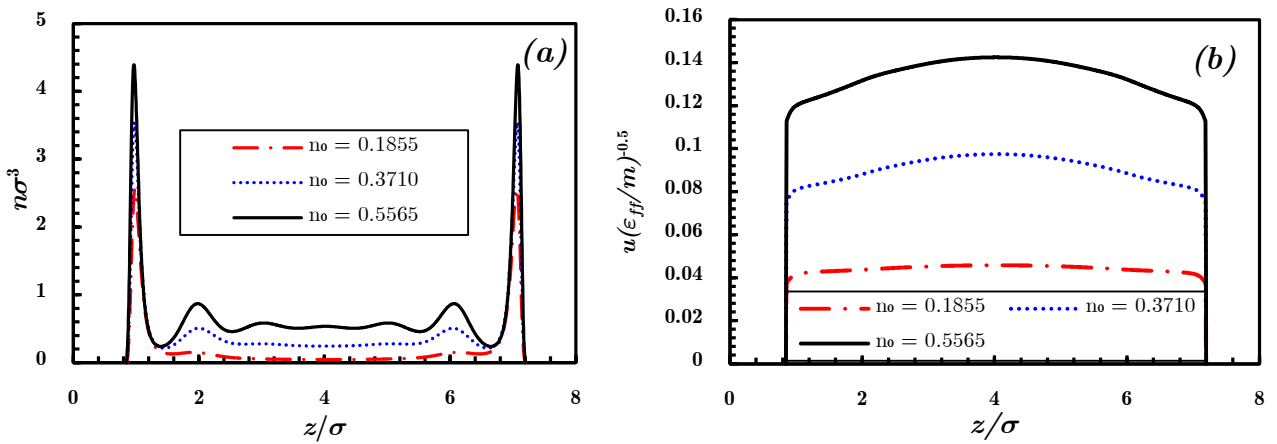

Figure 8: Density (*a*) and velocity (*b*) profiles of shale gas flow in organic nanopores under different pore averaged density conditions; $n = 0.1855\ \sigma^{-3}$, $n = 0.3710\ \sigma^{-3}$ and $n = 0.5565\ \sigma^{-3}$ correspond to 95 kg/m$^3$, 190 kg/m$^3$ and 285kg/m³ in SI units, which represent the fluid at the pressure of 16.7 MPa, 37.9 MPa and 83.2 MPa respectively. The temperature is kept at 363 K.

### 3.2 Effects of pore size

With the increase of the pore size, the number and the location of the adsorption layers are barely changed, while the bulk region becomes larger, as we can see from Figure 9a. This means that the adsorbed gas proportion is larger in small pores, and the free gas becomes predominant at large pores. The overall and slip flow velocity increase with pore size, indicating that the macro pores are still the most effective flow paths in unconventional shale gas reservoirs, while the micro pores storage the gas.

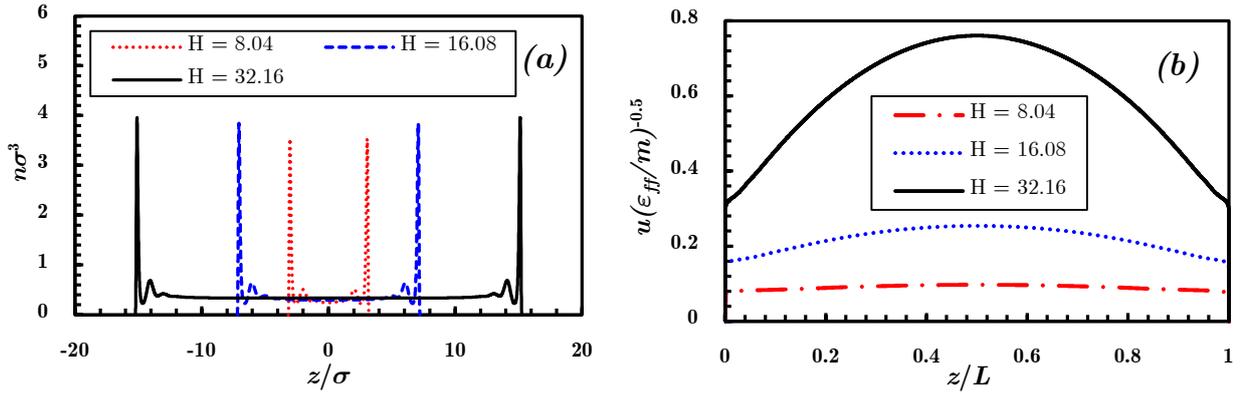

Figure 9: The effects of pore size on density (*a*) and velocity (*b*) profiles of shale gas flow.

### 3.3 Effects of temperature

The temperature range considered in the present work has insignificant influence on the cross-sectional density distribution of methane, as shown in Figure 10a. With the increase of the temperature, a certain amount of adsorbed gas molecules escapes from the adsorption layer into the bulk phase, due to the increase of the fluid kinetic energy. Meanwhile, methane molecules move faster under high temperature conditions, as the viscosity decreases with the increasing temperature, which helps the extraction of shale gas.

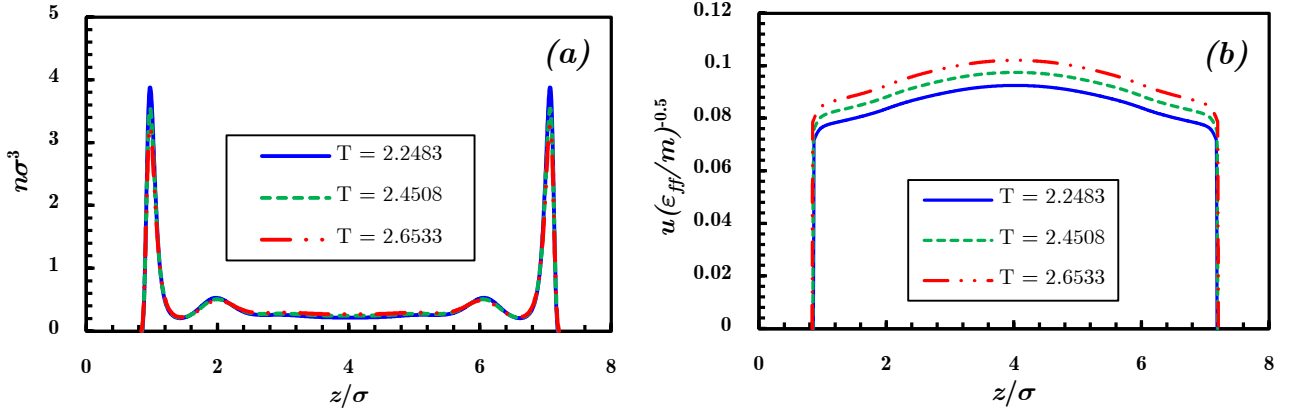

Figure 10: The effects of temperature on density ($a$) and velocity ($b$) profiles of shale gas flow in organic nanopores. The temperature of $T = 2.2488\ \varepsilon_{ff}/k_B$, $T = 2.4508\ \varepsilon_{ff}/k_B$, and $T = 2.6533\ \varepsilon_{ff}/k_B$ correspond to $T = 333$ K, $T = 363$ K, and $T = 393$ K, respectively.

### 3.4 Effects of shale composition

Different shale composition is characterized by different value of energy parameter of the surface $\varepsilon_{ww}$ in our study. Generally, the energy parameter of organic kerogen is larger than that of inorganic quartz. Therefore, the more organic matters in shale rocks, the larger the energy parameter $\varepsilon_{ww}$ would be. The energy parameters (gas-gas and gas-solid) relates the wettability of fluid on rock surface by (Barrat & Bocquet 1999)

$$\cos\theta = -1 + 2\frac{n_s \varepsilon_{wf}}{n_f \varepsilon_{ff}}, \qquad (25)$$

where $\theta$ is the contact angle, and $n_f$ is the gas density of the first adsorbed layer.

Meanwhile, the friction coefficient $\zeta$ is also related to the wettability in some relationship (Blake 1990). For a more general expression, the slip velocity $u_s$ in Eq.(24) reads

$$u_s = H\frac{G_x}{2\zeta_0}n_0\delta\left[e^{C(1-\cos\theta)/k_B T} - 1\right] \qquad (26)$$

where $\delta$ is the average distance between the centers of adjacent molecules, and $C$ is a coefficient relating to fluid-solid properties, which can be determined by experiment (Blake

1990).

As we can see from Eq.(25), the ratio of methane-rock energy parameter to methane-methane parameter, defined as $\gamma = \varepsilon_{wf}/\varepsilon_{ff}$, is a predominant factor that controls the wettability and the formation of adsorption layers in shale nanopores. With the decreasing of the ratio $\gamma$, the methane wettability on rock surface becomes weaker, and the rock characteristics tends from organic to inorganic materials. In this study, the rock energy parameter $\varepsilon_{ww}$ is adjusted to obtain different values of the ratio $\gamma$, as in Figure 11. As we can see from Figure 11(a), the adsorbed gas exists in both organic kerogen and inorganic minerals. Methane molecules accumulate more near the wall under stronger wettability condition, and consequently resulting a lower bulk density at the center. In Figure 11(b), the velocity profiles across pores with different shale composition are displayed. The slip velocity increases significantly with the decreasing wettability in an exponential form. Therefore, gas molecules moves much faster in inorganic pores than in kerogen. The surface diffusion flux is controlled simultaneously by adsorbed gas density and its moving velocity. Therefore, the contribution of surface diffusion in inorganic pores may also be significant due to the large moving velocity, which was usually ignored in previous studies due to the low adsorbed gas density. The surface diffusion contribution needs systematically study in the future.

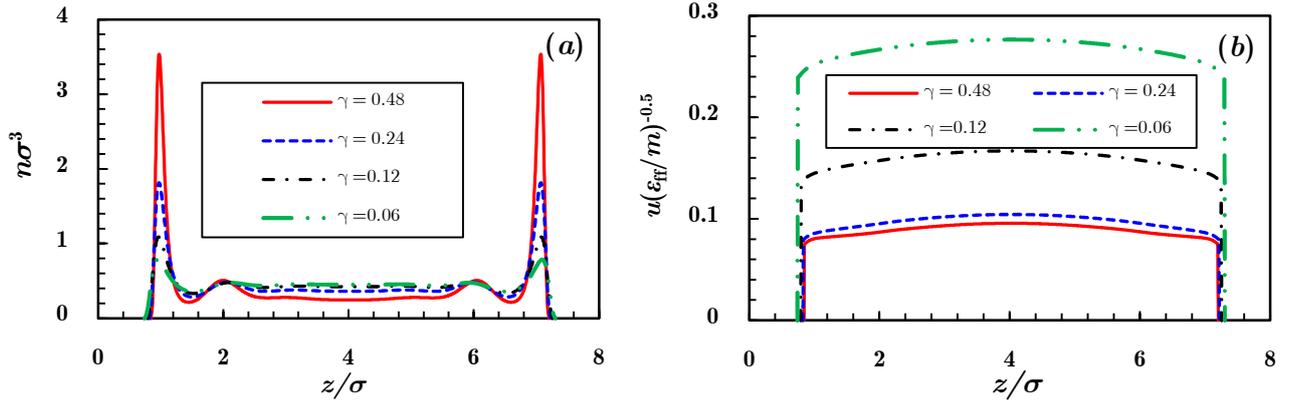

Figure 11: The effects of shale composition on gas dynamics in nanopores: ($a$) density and ($b$) velocity. The parameter $\gamma$ is defined as $\gamma = \varepsilon_{wf}/\varepsilon_{ff}$, which represents the strength of the competition between gas-surface and gas-gas molecules. The stronger the value $\gamma$, the stronger the surface action on methane molecules, and so as to the stronger wettability.

### 3.5 Comparison and analysis

The viscous flow, slip flow, Knudsen diffusion and surface diffusion are usually believed to play important roles in shale gas transport at the pore scale (Zhang et al. 2019). In the past decade, significant efforts have been made to unravel underlying flow physics, e.g. (Zhang et al. 2019; Javadpour et al. 2007; Wu et al. 2016a; Cai et al. 2019a; Darabi et al. 2012). Various approaches to extend the conventional Darcy's law to obtain apparent permeability, which can fit the experimental data well, have been reported. Among these models, a common one is to use a second-order slip boundary condition (Zhang et al. 2019). In our model, the flow of adsorbed gas and free gas are modelled in a self-consistent way, where the formation of adsorption layers is a result of the competition between rock-methane and methane-methane interactions. The non-dimensional velocity profiles in nano-slits with the separation of $H = 8.04\ \sigma$ and $H = 32.16\ \sigma$ are compared with the analytical solution of the N-S equation using the second-order slip boundary condition. Comparing to the parabolic velocity of the N-S solution, the velocity of shale gas flow in nanopores is more plug-like. This is caused by the large effective in highly-confined nanopores. With the increase of the pore size, the plug-like velocity tends to parabolic gradually. Although the second-order slip boundary condition takes the surface-slip effect into account, the slip velocity is much smaller than that predicted from the kinetic model. Generally, the dense gas flow in highly confined pores is significantly different from that predicted from the N-S or the Boltzmann equation.

The enhancement factors $\beta$ for $H = 8.04\ \sigma$ and $H = 32.16\ \sigma$ can be seen in Table 7. The flow in small pores is enhanced when $H = 8.04\ \sigma$, while it is hindered by the real gas effect and the pore confinement when the pore size is greater than some critical value. The N-S prediction underestimates the flow flux by 50.08% when $H = 8.04\ \sigma$, while it overestimates the flow flux by 34.16% when the pore size increases to $H = 32.16\ \sigma$.

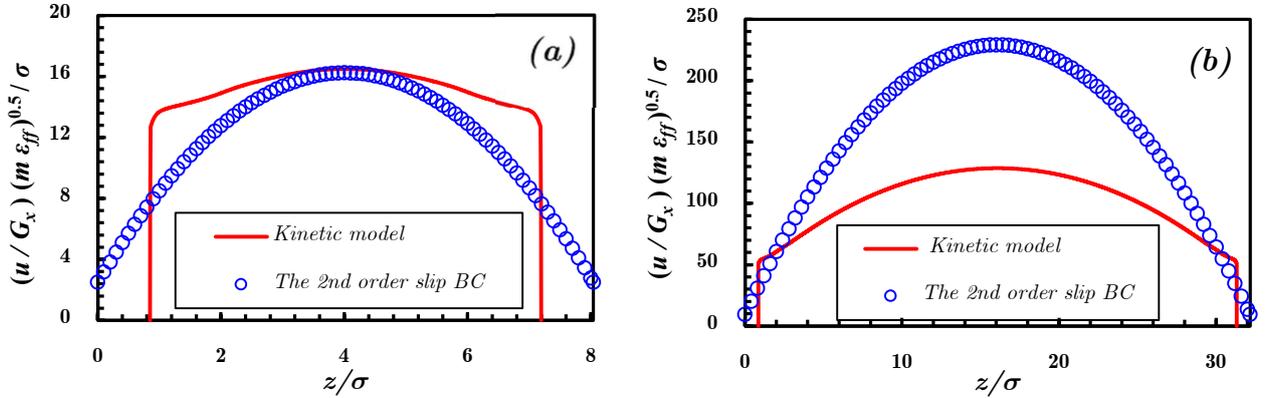

Figure 12: The comparison of velocity profiles from the kinetic model and the Navier-Stokes solution under the second-order boundary condition, which can be expressed as $u_s - u_w = \pm C_1 \lambda \left( \partial u / \partial n \right)_s - C_2 \lambda^2 \left( \partial^2 u / \partial n^2 \right)_s$. A systematic introduction about the second-order boundary condition can be referred to Zhang et al. (Zhang et al. 2019). In this paper, the coefficient $C_1$ and $C_2$ are taken from Chapman and Cowling (Chapman & Cowling 1970) as $C_1 = 1.0$ and $C_2 = 0.5$. The viscosity is evaluated at the pore averaged density according to Eq.(19) in the N-S calculations, which ignores the viscosity inhomogeneity.

Table 7: Comparison of enhancement factor from the results of the kinetic equation and the N-S equation with a second-order slip boundary condition. $\beta$ is the enhancement factor defined as $\beta = Q_{app} / Q_{ins}$.

| $H$ | $Kn$ | The kinetic model $\beta_k$ | The N-S equations $\beta_s$ |
|---|---|---|---|
| 8.04 $\sigma$ | 0.0437 | 1.7746 | 1.2738 |
| 32.16 $\sigma$ | 0.0109 | 0.7247 | 1.0663 |

**4 Conclusions**

Accurate modeling of non-equilibrium effect, dense gas effect and pore confinement effect is of significantly importance to understand shale gas transportation at the pore scale. In this paper, a self-consistent model is established, where the finite size of gas molecules and the non-local collisions are considered by the generalized Enskog equation. Gas molecular interaction is projected into the short-repulsive part and long-attractive part, which are modeled by the hard-sphere potential and the mean-field theory, respectively. The confinement effect is considered in the model by exerting a 10 – 4 – 3 LJ potential on fluid/surface molecules.

In highly-confined pore space, the dynamics of the dense gas and dilute gas are quite different even with the same Knudsen number. The dense gas usually displays plug-like velocity profile due to the high viscosity under confinement, while the dilute gas always produces a parabolic-like velocity. The Klinkenberg effect, which can describe slippage effect at low pressure conditions, cannot be applicable for dense gases under high pressure conditions. The adsorption layers exist in both organic and inorganic pores and shale gas in inorganic pores moves much faster than that in organic kerogen. Consequently, the surface diffusion flux in inorganic pores may also play an important role in shale gas transport.